\documentclass{article}
\usepackage{lscape,epsfig}
\usepackage{graphicx}
\usepackage{amssymb}
\usepackage{amsmath}
\usepackage{natbib}
\textheight=22cm \DeclareSymbolFont{ppa}{OT1}{ppl}{m}{it}
\DeclareMathSymbol{\vv}{\mathalpha}{ppa}{'166}

minus 2.3mu \thickmuskip = 2.6mu plus 2mu minus 2.6mu

\begin{document}

\newcommand{\dd}{\,{\rm d}}
\newcommand{\ie}{{\it i.e.},\,}
\newcommand{\etal}{{\it $et$ $al$.\ }}
\newcommand{\eg}{{\it e.g.},\,}
\newcommand{\cf}{{\it cf.\ }}
\newcommand{\vs}{{\it vs.\ }}
\newcommand{\zdot}{\makebox[0pt][l]{.}}
\newcommand{\up}[1]{\ifmmode^{\rm #1}\else$^{\rm #1}$\fi}
\newcommand{\dn}[1]{\ifmmode_{\rm #1}\else$_{\rm #1}$\fi}
\newcommand{\upd}{\up{d}}
\newcommand{\uph}{\up{h}}
\newcommand{\upm}{\up{m}}
\newcommand{\ups}{\up{s}}
\newcommand{\arcd}{\ifmmode^{\circ}\else$^{\circ}$\fi}
\newcommand{\arcm}{\ifmmode{'}\else$'$\fi}
\newcommand{\arcs}{\ifmmode{''}\else$''$\fi}
\newcommand{\MS}{{\rm M}\ifmmode_{\odot}\else$_{\odot}$\fi}
\newcommand{\RS}{{\rm R}\ifmmode_{\odot}\else$_{\odot}$\fi}
\newcommand{\LS}{{\rm L}\ifmmode_{\odot}\else$_{\odot}$\fi}

\newcommand{\Abstract}[2]{{\footnotesize\begin{center}ABSTRACT\end{center}
\vspace{1mm}\par#1\par \noindent {~}{\it #2}}}

\newcommand{\TabCap}[2]{\begin{center}\parbox[t]{#1}{\begin{center}
  \small {\spaceskip 2pt plus 1pt minus 1pt T a b l e}
  \refstepcounter{table}\thetable \\[2mm]
  \footnotesize #2 \end{center}}\end{center}}

\newcommand{\TableSep}[2]{\begin{table}[p]\vspace{#1}
\TabCap{#2}\end{table}}

\newcommand{\FigCap}[1]{\footnotesize\par\noindent Fig.\  %
  \refstepcounter{figure}\thefigure. #1\par}

\newcommand{\TableFont}{\footnotesize}
\newcommand{\TableFontIt}{\ttit}
\newcommand{\SetTableFont}[1]{\renewcommand{\TableFont}{#1}}
\newcommand{\MakeTable}[4]{\begin{table}[htb]\TabCap{#2}{#3}
  \begin{center} \TableFont \begin{tabular}{#1} #4
  \end{tabular}\end{center}\end{table}}

\newcommand{\MakeTableSep}[4]{\begin{table}[p]\TabCap{#2}{#3}
  \begin{center} \TableFont \begin{tabular}{#1} #4
  \end{tabular}\end{center}\end{table}}

\newenvironment{references}%
{ \footnotesize \frenchspacing
\renewcommand{\thesection}{}
\renewcommand{\in}{{\rm in }}
\renewcommand{\AA}{Astron.\ Astrophys.}
\newcommand{\AAS}{Astron.~Astrophys.~Suppl.~Ser.}
\newcommand{\ApJ}{Astrophys.\ J.}
\newcommand{\ApJS}{Astrophys.\ J.~Suppl.~Ser.}
\newcommand{\ApJL}{Astrophys.\ J.~Letters}
\newcommand{\AJ}{Astron.\ J.}
\newcommand{\IBVS}{IBVS}
\newcommand{\PASP}{P.A.S.P.}
\newcommand{\Acta}{Acta Astron.}
\newcommand{\MNRAS}{MNRAS}
\renewcommand{\and}{{\rm and }}
\section{{\rm REFERENCES}}
\sloppy \hyphenpenalty10000
\begin{list}{}{\leftmargin1cm\listparindent-1cm
\itemindent\listparindent\parsep0pt\itemsep0pt}}%
{\end{list}\vspace{2mm}}

\def\TYLDA{~}
\newlength{\DW}
\settowidth{\DW}{0}
\newcommand{\dw}{\hspace{\DW}}

\newcommand{\refitem}[5]{\item[]{#1} #2%
\def\REFARG{#3}\ifx\REFARG\TYLDA\else, {\it#3}\fi
\def\REFARG{#4}\ifx\REFARG\TYLDA\else, {\bf#4}\fi
\def\REFARG{#5}\ifx\REFARG\TYLDA\else, {#5}\fi.}

\newcommand{\Section}[1]{\section{#1}}
\newcommand{\Subsection}[1]{\subsection{#1}}
\newcommand{\Acknow}[1]{\par\vspace{5mm}{\bf Acknowledgements.} #1}
\pagestyle{myheadings}

\newfont{\bb}{ptmbi8t at 12pt}
\newcommand{\xrule}{\rule{0pt}{2.5ex}}
\newcommand{\xxrule}{\rule[-1.8ex]{0pt}{4.5ex}}
\def\thefootnote{\fnsymbol{footnote}}

\begin{center}
{\Large\bf
 The Clusters AgeS Experiment (CASE): \\
 The blue straggler Star M55-V60 caught amidst rapid mass exchange\footnote{This paper
 uses data obtained with the Magellan 6.5 m telescopes located at  Las Campanas Observatory, Chile.}}
 \vskip1cm
  {\large
        M.~~R~o~z~y~c~z~k~a$^1$,
      ~~J.~~K~a~l~u~z~n~y$^1$,
      ~~I.~B.~~T~h~o~m~p~s~o~n$^2$,
      ~~S.~~M.~~R~u~c~i~n~s~k~i$^3$,
      ~~W.~~P~y~c~h$^1$
      ~~and~~W.~~K~r~z~e~m~i~n~s~k~i$^1$
   }
  \vskip3mm
{ $^1$Nicolaus Copernicus Astronomical Center, ul. Bartycka 18, 00-716 Warsaw, Poland\\
     e-mail: (jka, mnr, wk, wp)@camk.edu.pl\\
  $^2$The Observatories of the Carnegie Institution of Washington, 813 Santa Barbara
      Street, Pasadena, CA 91101, USA\\
     e-mail: ian@obs.carnegiescience.edu\\
  $^3$Department of Astronomy and Astrophysics, University of Toronto,
     50 St. George Street, Toronto, ON M5S 3H4, Canada\\
     e-mail: rucinski@astro.utoronto.ca}
\end{center}

\vspace*{7pt} \Abstract {We analyze light and velocity curves of the
eclipsing blue straggler V60 in the field of the globular cluster
M55. We derive $M_p=1.259\pm 0.025\,M_\odot$, $R_p=1.102\pm
0.021\,R_\odot$, $M^{bol}_p=3.03\pm 0.09$ mag for the primary and
$M_s=0.327\pm 0.017\,M_\odot$, $R_s=1.480\pm 0.011\,R_\odot$,
$M^{bol}_s=4.18 \pm 0.12$ mag for the secondary. We measure an apparent
distance modulus $(m-M)_{V}=14.04\pm 0.09$ mag. Based on the systemic
velocity, distance, and proper motion of V60 we conclude that the
system is a member of the cluster and argue that its present state
is a result of  rapid but conservative mass exchange which the
binary is still undergoing. We report a peculiar blue excess on the
ascending branch of the primary eclipse of V60 and discuss its possible
origin. } {globular clusters: individual: M55 -- blue stragglers --
stars: individual: V60-M55
}

\Section{Introduction} The eclipsing variable M55-V60 (henceforth
V60) was discovered in 1997 within the CASE project (Kaluzny et al.
2010). Located 27 arcsec from the center of M55, it is a proper
motion member of the cluster (Zloczewski et al. 2011). At
$V_{max}=16.8$ mag and $(B-V)_{max}=0.41$ mag the system belongs to the
population of blue stragglers. Kaluzny et al. (2010) compiled
photometric data from several seasons and found that the light curve
was characteristic of short-period low-mass Algols: deep primary
eclipses with $DV\approx 1.8$~mag were followed by shallow secondary
ones with $DV\approx 0.20$~mag. The orbital period, equal to
1.183~d, was found to systematically increase. Such properties
indicate a semidetached binary in the mass-transfer phase, with the
donor being the less massive (secondary) component filling its Roche
lobe.

In this paper we analyze the velocity curve of V60 and refine the
preliminary photometric solution obtained by Kaluzny et al. (2010). Spectroscopic
and photometric data are described in Sect.~\ref{sec:photo} and
Sect.~\ref{sec:spectra}. The analysis of the data is detailed in
Sect.~\ref{sec:analysis} and our results are discussed in Sect.~\ref{sec:disc}.

\section{Photometric observations}
\label{sec:photo}

The photometric data collected between 1997 and 2009 were discussed in Kaluzny
et al. (2010), but for completeness we repeat here the most important points of
their discussion.

According to their findings, season-to-season variations of the
light curve in quadratures and in the secondary eclipse do not
exceed 0.010-0.015~mag in $V$-band. A significantly stronger
variability is observed in the primary minimum, whose depth ranged
from $V=18.62$~mag in 1999 to $V=18.57$ mag in 2007 and 2009. The
primary minimum is symmetric in $V$, allowing for a precise
determination of times of minimum (however, as we show and discuss
in Sect. \ref{sec:analysis}, the symmetry is broken in $B$). The $O-C$
diagram of the times of minimum (see Fig 3. in Kaluzny et al. 2010) 
indicates that the orbital period of V60, which in 2008 was equal to 
1.1830214$\pm$0.0000007~d, is increasing at a rate $dP/dt=3.0\times 10^{-9}$;
i.e. it would be doubled in just $\sim$10$^6$ years. 

A preliminary photometric solution favors a
semi-detached configuration with the secondary filling its Roche
lobe, consistent with the observed behavior of the period. The
luminosity ratios of the components in $V$ and $B$ bands (obtained
from the same solution at quadratures) yield apparent colors
$(B-V)_p \approx 0.3$ mag and $(B-V)_s \approx 0.9$ mag,
respectively for the primary and the secondary. With $V_p = 17.11$
mag the primary of V60 is located among the SX Phoenicis pulsating
variables on the color-magnitude diagram (CMD) of M55. A sinusoidal
variation with an amplitude of $\sim$0.005 mag and a period of
0.03087$\pm$0.00007 d is indeed observed in out-of eclipse sections
of the nightly $V$-light curves.

For the present analysis we use observations from five seasons in which both
eclipses were
covered (i.e. 1999, 2006, 2007, 2008 and 2009). The
data were phased according to the times of minima given in Table 4 of Kaluzny et
al. (2010) and merged into the composite $B$ and $V$ light curves (with 663 and
2666 points, respectively) shown in Fig.~\ref{fig:all_seasons_lc.eps}.

\section {Spectroscopic observations and orbital parameters of V60}
\label{sec:spectra}
Our radial velocity data are based on eight observations obtained with
the MIKE Echelle spectrograph (Bernstein et al. 2003) on the Magellan~II
(Clay) telescope at the Las Campanas Observatory. Seven of them were
made in 2004 between June 27th and October 3rd, and the eight one in 2005
on September 11th. Each observation consisted of two 1200--1800~s exposures
interlaced with an exposure of a Th/Ar lamp. For all observations a
$0.7\times 5.0$ arcsec slit was used, and $2\times 2$ pixel binning was applied.
At 438~nm the resolution was $\sim$2.7 pixels at a scale of 0.0043~nm/pixel.
The spectra were processed using a pipeline developed by Dan Kelson following
the formalism of Kelson (2003). In the blue channel of MIKE (380 -- 500 nm)
the average S/N ratio ranged between 21 and 25. As in the red channel (490 --
1000 nm) it was markedly worse, for radial velocity measurements data
from the blue channel only were used.

The original data had a high resolving power (about 40,000), but
because the S/N per pixel was rather low, we rebinned the
continuum-rectified spectra to a log($\lambda$) scale, convolved
them with a Gaussian with FWHM=15 km/s, and rebinned them to 3 km/s
steps (i.e. 5 pixels per a resolution interval or 2.5 times
over-sampling). This procedure improved the quality of the spectra
and adjusted the resolution to better reflect the actual rotational broadening of the
lines. The final spectra have  22,300 data points in the wavelength interval of
400 - 500 nm, where the S/N ratio was the highest. The data
reprocessed in this way were analyzed with the help of a code based on
the broadening function (BF) formalism of Rucinski (2002). The BFs
were determined over 261 steps of 3 km/s in a process which is
effectively a least squares solution of 22,300 linear equations with
a 22,300/261 = 85 times over-determinacy. Templates with $[{\rm
Fe/H}]=-2.0$, almost exactly matching the value of -1.94 given by
Harris (1996; 2010 edition), were selected from the Synthetic
Stellar Library of Coelho \etal (2005). Measurement errors,
estimated based on the residuals from the fit described in the next
paragraph, amount to 3.6~km~s$^{-1}$ (primary) and 1.9~km~s$^{-1}$
(secondary). Rotational profile fits which are performed
automatically within the BF formalism yielded $v_{rot} \sin i$ equal
to $46.1\pm1.4$~km~s$^{-1}$ for the primary and
$63.2\pm2.2$~km~s$^{-1}$ for the secondary (corrections for the
assumed Gaussian smoothing with FWHM = 15 km/s were taken into
account).

The velocity curve was fitted with the help of the spectroscopic data solver
written and kindly provided by Guillermo Torres. On the input to the solver
the observed velocities had to be transformed into mass-center velocities of
the components in order to account for their asphericity. This was done
iteratively by applying the solver to the uncorrected velocities, feeding the
solution into PHOEBE31a implementation (Pr\v{s}a \& Zwitter 2005) of the
Wilson-Devinney model (Wilson \& Devinney 1971; Wilson 1979), finding
the corrections, and applying the solver again to the corrected velocities.
The advantage of this
procedure is that Torres's code automatically calculates the errors of the
fitted orbital parameters, which otherwise would have to be estimated with PHOEBE
by Monte-Carlo techniques. The measured orbital velocities are listed in
Table~\ref{tab:rv} together with mass-center corrections and residuals from the
fit. The fitted velocity curve generated by PHOEBE is shown in Fig.~\ref{fig:rv},
and the orbital parameters obtained from the fit are listed in Table \ref{tab:orb_parm}
together with formal 1-$\sigma$ errors returned by the fitting routine.

\section {Photometric solution and system parameters}
\label{sec:analysis}

\begin{figure}
 \centerline{\includegraphics[width=0.95\textwidth,
             bb= 35 334 564 690,clip ]{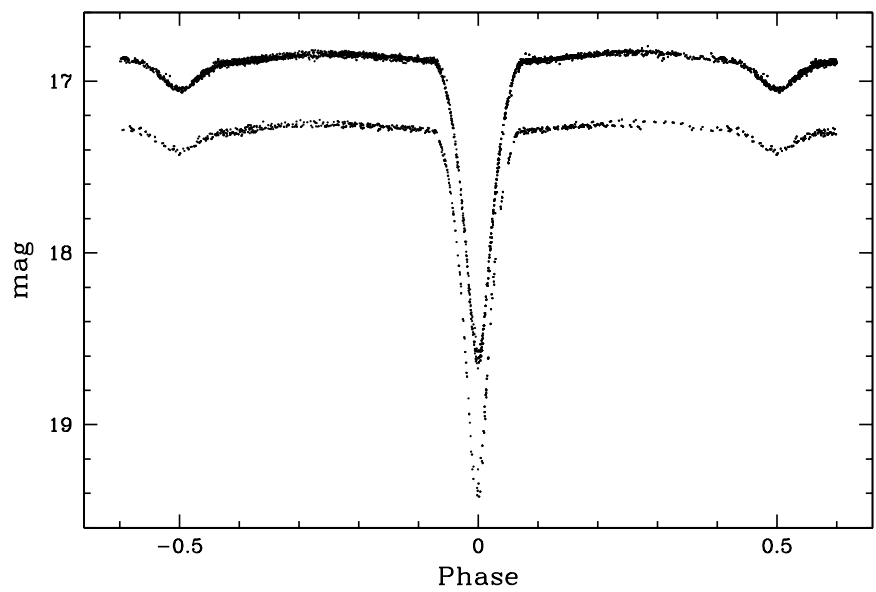}}
 \caption
  {Light curve of V60 in bands $V$ (upper) and $B$ (lower). Data from five observing
   seasons are shown.
   \label{fig:all_seasons_lc.eps}
  }
\end{figure}

\begin{figure}
 \centerline{\includegraphics[width=0.95\textwidth, bb= 19 285 564 690,clip ]{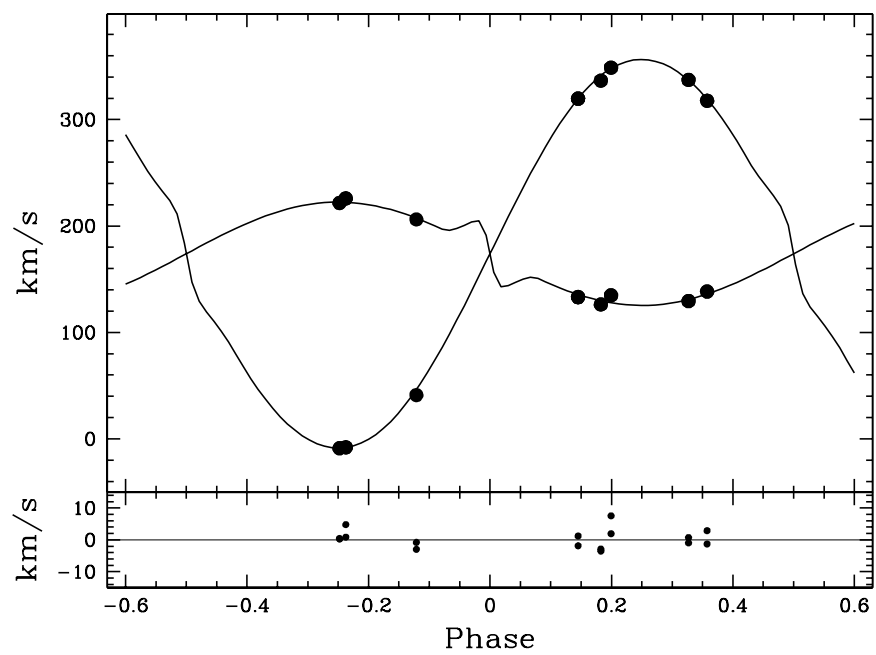}}
 \caption
  {Radial velocity curve of V60 (up er panel), and residuals from the fit (lower panel).
           \label{fig:rv}
  }
\end{figure}

\begin{figure}
 \centerline{\includegraphics[width=0.95\textwidth,
             bb= 18 438 564 690,clip ]{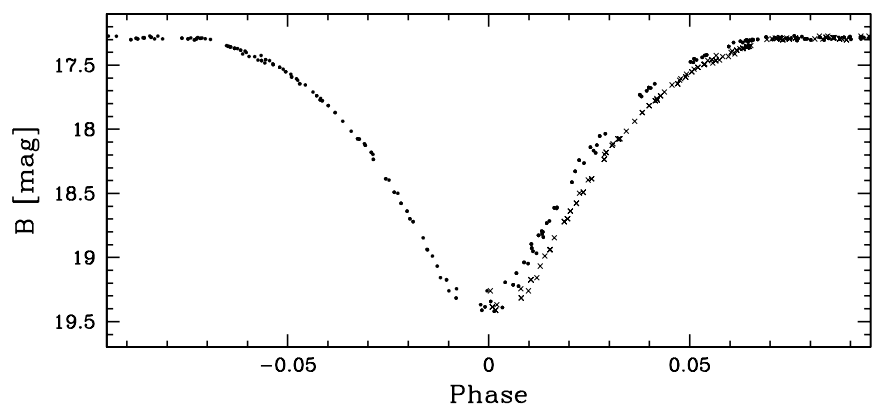}}
 \caption
  {Asymmetry of the primary eclipse of V60 in the $B$-band. Points: observational
   data from five observing seasons phased as explained in Sect. \ref{sec:photo}.
   Crosses: descending branch reflected in phase.
           \label{fig:Bmain_asymmetry}
  }
\end{figure}

A closer look at the primary eclipse in $B$ reveals a significant
asymmetry identifiable in each observing season: the ascending
branch is by up to 0.2~mag bluer than the descending one (see
Fig.~\ref{fig:Bmain_asymmetry}). As such an excess is not possible
to model within the standard approach, we were forced to discard the
affected observational points between phases 0.0 and 0.07. In
principle, the pulsational modulation with an amplitude
of$\sim$0.005 mag in $V$ mentioned in Sect. \ref{sec:photo}, which
is at least partly responsible for the vertical scatter of points in
Fig.~\ref{fig:all_seasons_lc.eps} should also be removed. However,
an amplitude this low is suggestive of nonradial pulsations which
are commonly observed in SX Phe stars (e.g. Pych et al. 2001; Olech
et al. 2005), and the results of such a procedure applied to the
primary eclipse would be unreliable.

The photometric solution was also found with the help of PHOEBE
utility which employs the Roche geometry to approximate the shapes
of the stars, uses Kurucz model atmospheres, treats reflection
effects in detail, and, most importantly, allows for the
simultaneous analysis of $B$ and $V$ data. For the radiative envelope of
the primary we adopted a gravity brightening coefficient $g_p = 1.0$
and a bolometric albedo $A_p = 1.0$. The same coefficients for the
convective envelope of the secondary were set to $g_s = 0.32$ and
$A_s = 0.5$. The effects of reflection were included.
Limb darkening coefficients were
interpolated from the tables of Claret (2000) with the help of the
JKTLD code.\footnote{Written by John Southworth and available at
www.astro.keele.ac.uk/jkt/codes/jktld.html}
Full synchronization of both components was assumed.

An approximate temperature of the primary $T_p$ was calculated from
the dereddened $B-V$ index obtained by Kaluzny et al. (2010), using
a color-temperature calibration based on the data from the Dartmouth
Stellar Evolution Database (Dotter et al. 2008). We decided to
employ the synthetic calibration because the starting value
$(B-V)_{p0}=0.23$ mag was too close to the applicability limit of the
empirical calibration compiled by Casagrande et al. (2010) which is
formally valid for 0.18 mag $<(B-V)<$ 1.29 mag (this issue is further
discussed in the last paragraph of this Section). For dereddening a
value $E(B-V)=0.08$ mag with an assumed error of 0.01 mag was used (Harris
1996, 2010 edition). The next approximation was found based on
PHOEBE-provided contributions of each component to the total light
at quadratures in $B$ and $V$ bands which allow to calculate the
updated observed $(B-V)_p$. The updated $(B-V)_p$ was dereddened and
translated into temperature the same way as before. The procedure
was repeated until the convergence was reached at $T_p=8160\pm140$~K
(the error is due to uncertainties in calibration, reddening and
zero points of $B$ and $V$ photometry). The temperature of the
secondary, $T_s$, was automatically adjusted by PHOEBE.

\begin{figure}
 \centerline{\includegraphics[width=0.95\textwidth,
             bb= 22 409 564 690,clip ]{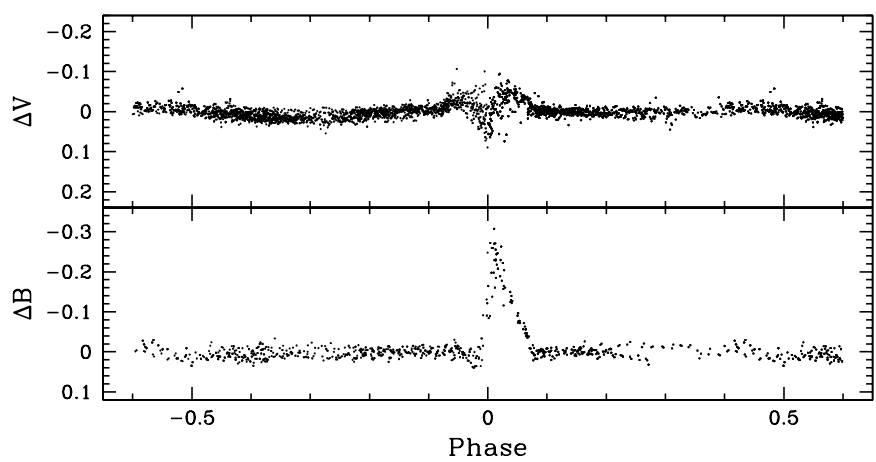}}
 \caption
  {Residuals from the fits to $V$ and $B$ light curves. The $B$-residuals include
   points on the ascending branch of the primary eclipse which were omitted during
   the fitting (see text for explanations). The vertical scale is the same in both
   panels.
           \label{fig:VB_res.eps}
  }
\end{figure}

The fitted photometric parameters of V60 are listed in
Table~\ref{tab:phot_parm} together with the errors estimated based
on additional fits to each of the seasonal $B$ and $V$ light curves
(the ascending branch of the primary eclipse was removed from each
$B$-curve while fitting). Table~\ref{tab:abs_parm} contains the
final absolute parameters of the system, hereafter referred to as
the standard solution. The residuals from the final fit are shown in
Fig.~\ref{fig:VB_res.eps}. The blue excess on the ascending branch
of the main eclipse is clearly visible; it is also evident that the
largest $V$-residuals occur within the main eclipse. Apparently, the
stream(s) of gas between the components generate an additional light
in the system which PHOEBE is not able to account for. The likely
variability of such a light source might be responsible for the
enhanced scatter observed in $V$-band at the bottom of the main
eclipse.

For $T_p=8160$ K and the primary's gravitational acceleration
$g_p=4.4$ obtained from our solution, the bolometric correction
amounts to -0.07 mag (Dotter et al. 2008; on-line version of
Dartmouth Stellar Evolution Database). With $M_{bol}=3.03 \pm
0.09$~mag (Table 4), the absolute $V$ magnitude of the primary is
$M_{Vp}=3.10\pm0.09$ mag. From the light curve solution we obtained
$V_{p}=17.144\pm0.014$ mag, where the error includes uncertainty of the
zero point of the photometry. The apparent distance modulus of V60
is then $(m-M)_{V}=14.04\pm 0.09$ mag. This is consistent with the value
of 13.89 listed by Harris (1996; 2010 edition). With $E(B-V)=0.08
\pm 0.01$ mag, the foreground absorption in $V$-band is $A_{V}=0.25\pm
0.03$ mag, corresponding to an absolute distance modulus
$(m-M)_{0}=13.79\pm0.10$ mag. A very similar value follows from the
result obtained by isochrone fitting (Dotter et. al. 2010), who
derived an apparent distance modulus of 13.88 mag for the ACS/F814W
filter. Since from the calibration of Girardi et al. (2008) one gets
$A_{F814W}=0.15$ mag, the absolute distance modulus becomes
$(m-M)_{0}=13.73$ mag.

We would like to note, however, that the primary's bolometric
luminosity we derive strongly depends on the effective temperature
deduced from the color index. If instead of $T_p=8160$ K we used
$T_p=7450$ K resulting from the empirical calibration of Casagrande
et al. (2010), we would obtain an unacceptably low value of the
absolute distance modulus $(m-M)=13.39$ mag. This casts come doubts on
the validity of that calibration for blue low-metallicity stars, and
indeed, a closer inspection of Fig. 14 in Casagrande et al. (2010)
strongly suggests that for $[{\rm Fe/H}]=-2.0$ and $B-V<0.35$ mag 
their $T_{eff}-(B-V)$ relation is but an extrapolation. 

\Section{Discussion}
 \label{sec:disc}
 The important conclusion
following from our analysis is that the mass transfer in V60 must be
to a good approximation conservative. This is because the total mass
of the system, $M=1.59$~$\mathrm{M_\odot}$, is almost exactly two
times larger than the M55 turnoff mass of
$\sim$0.8~$\mathrm{M_\odot}$ (e.g. Zaggia et al. 1997). This,
together with the rate of period lengthening and system parameters
found in Sect.~\ref{sec:analysis}, implies an orbital expansion rate
$dA/dt = 3.4\times10^{-6}$~$\mathrm{R_\odot y^{-1}}$ and a mass
transfer rate $dM/dt = 1.4\times10^{-7}$~$\mathrm{M_\odot y^{-1}}$.
Thus, V60 is in a phase of a rapid mass exchange, which, given the
low mass of the H-shell burning secondary, cannot last longer than a
few hundred thousand years. With this in mind, it is interesting to
see that the present primary must be quite far from thermal
equilibrium, as its temperature is over a thousand K lower than the
temperature of a $1.1$~$\mathrm{R_\odot}$ star on the
$1.26$~$\mathrm{M_\odot}$ evolutionary track, which according to
Dotter et al. (2008) should be equal to 9350~K. We also note that the
original mass ratio must have been close to unity -- otherwise the
original primary would have left the main sequence much earlier,
and the system would be in a more advanced evolutionary state (e.g. detached,
or semidetached with the present primary filling its Roche lobe).

The blue excess of the system between phases 0.0 and 0.07 (hereafter BEx) is a
puzzling feature which to our knowledge has no counterpart in other
Algols. Suspecting it might be spurious, we checked if there was a systematic
shift between phases calculated for $B$ and $V$ data points. None was detected.
We then tried various means to remove BEx or at least to make it weaker. First,
we fitted the complete $B$ curve allowing PHOEBE to find a global phase shift
$\Delta\varphi$ that would center the main $B$-minimum at phase 0. Subsequently,
with $\Delta\varphi$ kept fixed, both $B$ and $V$ curves were fitted. This
procedure indeed made the blue excess smaller, but the $V$-fit became significantly
worse both in the primary and the secondary minimum. In a second series of
experiments we fitted the curves with the {\em descending} branch removed from
either $B$ or $B$ and $V$ primary minima, allowing for a global phase shift or
keeping it fixed at 0. Since the residuals did not look any better in any of those
fits, we had to accept the standard solution as the best one we were able
to obtain. The encouraging finding was that the parameters of all the trial fits
did not differ from the standard ones by more than a factor of $\sim$1.5 times the errors given
in Table~\ref{tab:abs_parm}. In particular, this means that the masses of the
components are determined sufficiently accurately for the conclusion concerning the
conservative mass transfer to be firmly footed. We note, however, that while the
mass exchange was conservative or nearly conservative, the system must have lost an
appreciable amount of the orbital angular momentum $J_{orb}$, possibly by magnetic
braking (see e.g. Eggleton and Kiseleva-Eggleton 2002). At the present value of
$J_{orb}$ the minimum orbital separation (which is achieved when $M_s=M_p$) would
be 0.94 $\mathrm{R_\odot}$  -- far too small to accommodate two 0.8 $\mathrm{M_\odot}$
stars. Of course, the loss of $J_{orb}$ must have occurred after the
equalization of masses, i.e. when the original mass ratio was reversed and the
orbital separation began to increase.

\begin{figure}
 \centerline{\includegraphics[width=0.53\textwidth, angle=270,
             bb= 135 90 437 637,clip ]{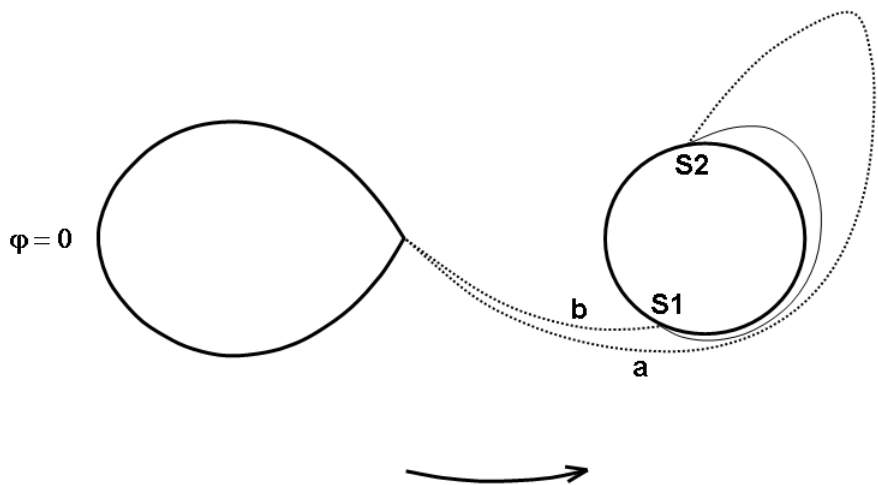}}
 \caption
  {Schematic view of V60 with sizes of the stars and binary separation drawn to
   scale. See text for explanations.
   \label{fig:sys}
  }
\end{figure}

Numerous spectroscopic and photometric effects observed in
Algol-like systems are attributed to three dynamical agents: a stream
of matter flowing out of the secondary, a regular or transient
accretion disk formed by the stream around the primary, and shock
waves excited where the stream hits the disk or directly impacts the
primary. V60 is a compact system in which there is no space for a
regular accretion disc (e.g. Richards 1992), so that the stream
directly hits the primary and induces shock waves(s) in its
atmosphere. BEx, if real, must also originate in a shock. The
problem is that the shock cannot be localized in the standard place
where the trailing hemisphere of the secondary is hit by the stream
(S1 in Fig.~\ref{fig:sys}): if that were the case, then BEx would
show up on the {\em descending} branch of the main minimum. We
calculate that if a particle leaving the L$_\mathrm{1}$ point is to miss
the trailing hemisphere and land onto the leading one (path a ending
at S2 in Fig.~\ref{fig:sys}) an initial velocity of
$\sim$0.5$v_{orb}$ is needed, where $v_{orb}$ is the orbital velocity
of the secondary around the mass center. Since in reality the
initial velocity is on the order of the thermal velocity, i.e. about a
tenth of $v_{orb}$, the stream must follow path b in
Fig.~\ref{fig:sys} and hit the star at S1. Another shock at S2 can
only be formed by the matter reflected off the primary and flowing
along the continuous line in Fig.~\ref{fig:sys}. Such an interpretation is
not a new one -- in fact, it closely resembles the schematic Algol
model derived from spectroscopic observations by Gillet et al.
(1989) and shown in their Fig. 3. Unfortunately, V60 is too faint for
a detailed spectroscopic study, and the only observational proof for
the existence of S2 would be a UV emission coinciding in phase with
BEx.

We note that speculations about the possibility of stream reflection are not entirely
unrealistic: a partial reflection was observed in models of flows in a cataclysmic
binary obtained by Rozyczka (1987). Admittedly, in those simulations the stream
reflected off the accretion disk, but the physics involved was sufficiently simple that the
results may be safely applied to a stellar atmosphere. It should also be mentioned
that a reflection from the primary component in an Algol-type system was observed
in simulations performed by Richards (1998), however it was to a large extent
predetermined by boundary conditions.

Assuming the above explanation of the origin of BEx is feasible we still have to
explain why the emission from S2 is only visible in a narrow range of phases, and
why the emission from S1 is not visible at all. The only possibility we see is
that both shocks are hidden behind gas stream(s): S1 entirely, and S2 partly. This
speculation can be verified by detailed hydrodynamical simulations which are beyond
the scope of this paper.

\Acknow
{We thank Joe Smak for helpful discussions and for the permission to use his
 code for integration of particle trajectories. JK and MR were supported by
 the grant 2012/05/B/ST9/03931 from the Polish National Science Centre.
}

\clearpage

\begin{table}
 \caption{Radial velocity observations of V60. \label{tab:rv}}
 \begin{tabular}{@{}lrrrrrrc}
 \hline
 \hline
HJD      & Phase & $v_{p}$   &  $v_{s}$ &  $dv_{p}$ & $dv_{s}$ & $(O-C)_{p}$ & $(O-C)_{s}$\\
-2453000 &       & [km/s]    &  [km/s]  &   [km/s] & [km/s] & [km/s]    &  [km/s]\\
\hline
183.79260 & 0.7523 & 221.66 &   -8.87 &  0.07 & -3.49 &  0.27 &  0.44 \\
206.73462 & 0.1451 & 133.18 &  319.61 & -0.07 &  1.72 & -1.84 &  1.21 \\
272.53145 & 0.7626 & 226.01 &   -7.92 &  0.07 & -3.46 &  4.77 &  0.85 \\
275.56501 & 0.3269 & 129.41 &  337.27 & -0.03 &  1.23 & -0.98 &  0.70 \\
276.57721 & 0.1825 & 126.29 &  336.59 & -0.07 &  2.54 & -2.88 & -3.51 \\
281.51625 & 0.3574 & 138.33 &  317.70 & -0.02 & -0.66 &  2.89 & -1.28 \\
282.51218 & 0.1994 & 134.86 &  348.79 & -0.07 &  2.92 &  7.54 &  1.94 \\
989.57690 & 0.8788 & 206.18 &   41.17 &  0.06 & -1.34 & -0.78 & -2.95 \\
\hline
\end{tabular}\\
\rule{0 mm}{4 mm}Columns $dv_{p}$ and $dv_{s}$ contain corrections which have
 to be added to velocities $v_{p}$ and $v_{s}$ in order to transform them into
 mass-center velocities of the primary ($p$) and the secondary ($s$).
\end{table}
\begin{table}
 \caption{Orbital parameters of V60 \label{tab:orb_parm} }
 \begin{tabular}{lcc}
  \hline
   Parameter  & Unit &  Value \\
  \hline
     $P$        &d           &$1.1830214\pm0.0000007^\mathrm{a}$ \\
     $\gamma$   &km s$^{-1}$ &173.14$\pm$0.68 \\
     $K_p$      &km s$^{-1}$ & 48.32$\pm$1.59 \\
     $K_s$      &km s$^{-1}$ &185.96$\pm$0.87 \\
     $e$        &     & 0.0$^\mathrm{b}$ \\
     $\sigma_p$   &km s$^{-1}$ & 3.56    \\
     $\sigma_s$   &km s$^{-1}$ & 1.91    \\
     Derived quantities:        &                 &\\
     $A\sin i$    &$R_\odot$ & 5.476$\pm$0.046 \\
     $M_p\sin^3 i$  &$M_\odot$ & 1.251$\pm$0.025 \\
     $M_s\sin^3 i$  &$M_\odot$ & 0.325$\pm$0.017 \\
  \hline
 \end{tabular}\\
\rule{0 mm}{3 mm}
$^\mathrm{a}$Obtained from photometry\\
\rule{1 mm}{0 mm}$^\mathrm{b}$Assumed in fit
\end{table}

\clearpage

\begin{table}
 \caption{Photometric parameters of V60.
          \label{tab:phot_parm}
         }
 \begin{tabular}{lccc}
  \hline
   Parameter      & Unit &  Value   \\
  \hline
     $i$          & deg  & 86.4$\pm$0.5 \\
     $e$          &      &   0$^\mathrm{a}$ \\
     $T_p$        & K    & 8160$^\mathrm{b}$ \\
     $T_s$        & K    & 5400$\pm$77   \\
     $R_p$        & $R_\odot$ & 1.102$\pm$0.019 \\
     $R_s$        & $R_\odot$ & 1.480$^\mathrm{c}$ \\
     $(L_p)_V$    & \%   & 76.5$\pm$1.8 \\
     $(L_p)_B$    & \%   & 85.3$\pm$1.5 \\
     $\sigma_V$ &mmag& 18 \\
     $\sigma_B$ &mmag& 12 \\
  \hline
 \end{tabular}\\
\rule{0 mm}{3 mm}
$^\mathrm{a}$Assumed in fit\\
\rule{1 mm}{0 mm}$^\mathrm{b}$Obtained from the synthetic $(B-V)-T$ calibration
           based on the Dartmouth Stellar Evolution Database\\
\rule{1 mm}{0 mm}$^\mathrm{c}$For fixed $A$ = 5.487 $R_\odot$
\end{table}

\begin{table}
 \caption{Absolute parameters of V60
          \label{tab:abs_parm}
         }
 \begin{tabular}{lcc}
  \hline
   Parameter     & Unit & Value  \\
  \hline
     $P$   & d          &$1.1830214\pm0.0000007$ \\
     $A$   & $R_\odot$  & 5.487$\pm$0.047 \\
     $i$   &  deg       &  86.4$\pm$0.5    \\
     $e$   &            &   0$^\mathrm{a}$  \\
     $M_p$ & $M_\odot$  & 1.259$\pm$0.025  \\
     $M_s$ & $M_\odot$  & 0.327$\pm$0.017  \\
     $R_p$ & $R_\odot$  & 1.102$\pm$0.021$^\mathrm{b}$   \\
     $R_s$ & $R_\odot$  & 1.480$\pm$0.011$^\mathrm{b}$   \\
     $T_p$ & K          & 8160$\pm$140$^\mathrm{c}$      \\
     $T_s$ & K          & 5400$\pm$160$^\mathrm{d}$       \\
     $M^{bol}_p$ & mag  &3.03$\pm$0.09 \\
     $M^{bol}_s$ & mag  &4.18$\pm$0.12 \\
  \hline
 \end{tabular}\\
\rule{0 mm}{3 mm}
$^\mathrm{a}$Assumed in fit\\
\rule{1 mm}{0 mm}$^\mathrm{b}$Errors include uncertainty in $A$\\
\rule{1 mm}{0 mm}$^\mathrm{c}$Combined error of calibration, reddening
      and zero points of $B$ and $V$ photometry\\
\rule{1 mm}{0 mm}$^\mathrm{d}$Error includes uncertainty in $T_p$
\end{table}

\clearpage

\clearpage

\clearpage

\clearpage

\clearpage

\end{document}